\documentclass[fleqn,twoside]{article}
\usepackage{espcrc2}

\usepackage{graphicx}
\usepackage[figuresright]{rotating}

\newcommand{\AmS}{{\protect\the\textfont2
  A\kern-.1667em\lower.5ex\hbox{M}\kern-.125emS}}

\title{$N_f=1$ QCD simulation with improved gauge action at 
finite temperature}

\def\be{\begin{equation}}
\def\ee{\end{equation}}
\def\bea{\begin{eqnarray}}
\def\eea{\end{eqnarray}}

\author{Fumiyoshi Shoji\address{Information Media Center,
 Hiroshima University, Higashi-Hiroshima, Japan, 739-8526} and
Tetsuya Takaishi\address{Hiroshima University of Economics, Hiroshima, Japan, 731-0192}
}

\begin{document}
\begin{abstract}
We study the deconfinement transition of $N_f=1$ QCD 
by the hybrid Monte Carlo algorithm with Wilson fermions. 
We calculate the Polyakov loop, its susceptibility and  Binder cumulant and 
use the $\chi^2$ method to locate the phase transition point.
Our results are similar to  the previous results obtained by the multiboson algorithm.
\vspace{1pc}
\end{abstract}

\maketitle

\section{Introduction}

L\"uscher proposed the multiboson algorithm\cite{Luscher,multibosonorg} in which 
a polynomial approximation to the inverse of the fermion matrix is used.
This algorithm is considered to be an alternative one to simulate dynamical QCD.
Bori\c{c}i and de Forcrand\cite{BF} noticed that the polynomial approximation 
can be used to simulate odd flavour QCD.
Then the multiboson algorithm is used to study the deconfinement
transition in one flavour QCD\cite{multiboson}.
The original hybrid Monte Carlo (HMC) algorithm\cite{HMC} is limited for even flavour case.
However the polynomial approximation can be also applied for the HMC\cite{FAST},
which means that one can also simulate odd flavour QCD by the HMC\cite{odd}.
We investigate the phase structure of one flavour QCD 
by the HMC and check whether different algorithms give the same results. 
We calculate Polyakov loop, its susceptibility and  Binder cumulant
and use the $\chi^2$ method\cite{engels} to locate the precise position of the phase transition.
Mainly we present results from the Wilson gauge action,
and also give preliminary results from the DBW2 gauge action\cite{DBW2}.

\section{$N_f=1$ algorithm}

The partition function of  $N_f=1$ QCD is given by
$\displaystyle
Z=\int [dU] \det D \exp(-S_{g}[U]).
$
One can approximate the inverse of $D$ by a polynomial\cite{Luscher} as
$\displaystyle
1/D\approx P_m(D) \equiv \prod_{k=1}^{m}(D-z_k),
$
where $z_k$ are the roots of the polynomial $P_m(D)$. 
For the even degree $m$,
det$D$ can be rewritten as
\be
\mathrm{det} D= C_n \mathrm{det}(T^{\dagger}_n(D)T_n(D))^{-1},
\ee
where $T_n(D)\equiv \prod_{k=1}^{n}(D-Z_{2k-1})$, and $C_n\equiv \mathrm{det}
(DT_n^\dagger(D)T_n(D))^{-1}$ is a correction factor to the polynomial approximation.
det$(T_n^\dagger(D)T_n(D))$ can be expressed as,
\bea
&&\mathrm{det}(T_n^\dagger(D)T_n(D))^{-1}\nonumber\\
&\approx& \int[d\phi^\dagger][d\phi]
\mathrm{exp}(-\phi^\dagger T_n^\dagger(D)T_n(D)\phi).
\eea
Then one can define $N_f=1$ Hamiltonian for HMC algorithm as, 
\be
H^{N_f=1}=\frac{1}{2}\sum_iP_i^2+S_g[U]+\phi^\dagger T_n^\dagger(D)T_n(D)\phi.
\ee
where $P_i$ are momenta which are conjugate to link variables.
$\phi^\dagger T_n^\dagger(D)T_n(D)\phi$ is manifestly positive,
which is essential to the implementation of the HMC.  
The correction factor $C_n$ is used to make the algorithm exact\cite{odd}. 

\begin{table*}[t]
\begin{tabular}{c|cc|cc|cc}
\hline
$\kappa$&\multicolumn{2}{c}{ $8^3 \times 4$} &
    \multicolumn{2}{c}{$ 12^3 \times 4$} &
\multicolumn{2}{c}{$ 16^3 \times 4 $ }\\
\hline
 & $n$/Acc & traj. & $n$/Acc & traj.& $n$/Acc& traj. \\
\hline
0.05& 8/0.86 & 10K & 12/0.75 & 12K & 24/0.62 & 12K \\
\hline
0.10& 16/0.86 & 10K & 24/0.74 & 12K & 32/0.62 & 12K \\
\hline
0.12& 24/0.86 & 15K & 32/0.75 & 20K & 40/0.62 & 14K \\
\hline
0.14& 32/0.86 & 10K & 40/0.74 & 14K & 50/0.61 & 12K \\
\hline
\end{tabular}
\caption{
$n$ and Acc are the number of degrees of polynomial
and acceptance ratio of the HMC respectively.
The step size $d\tau$ is set to $0.05$ for all simulations.
To obtain the best performance of the HMC one should tune $d\tau$
so that the acceptance is $60-70\%$\cite{HOHMC}.
In this study we have not done such tuning.
}
\label{table:parameters}
\vspace{-0.5cm}
\end{table*}

\section{$\chi^2$--method}


In this study we use $\chi^2$--method\cite{engels} to find critical $\beta$ for each $\kappa$.
By differentiating the singular part of the free energy density on 
$N_s^3\times N_t$ lattice 
and expanding these equations with respect to $x$,
we obtain the leading $N_s$ behavior at $x=0$,
\begin{eqnarray}
  \ln \langle L\rangle(x=0,N_s)&=&c_0\ln N_s + c_1
\label{eqn:Lfit} ,\\
  \ln \chi(x=0,N_s)&=&c_0\ln N_s + c_1 \label{eqn:chifit},\\
B_4(x=0,N_s)&=& c_0+c_1N_s^{-\omega}. \label{eqn:grfit}
\end{eqnarray}
where $\langle L \rangle$, $\chi$, $B_4$ and $x$ are
Polyakov loop, its susceptibility, Binder cumulant\cite{binder} and 
reduced temperature $(T-T_c)/T_c$ respectively:
\begin{eqnarray}
  L&=&\frac{1}{N_s^3}\sum_x \frac{1}{N}\mbox{Tr}\prod_t U_4(x,t)  \\
  \chi&=&N_s^3(\langle L^2\rangle - \langle L\rangle^2) ,
~~~\chi_v=N_s^3\langle L^2\rangle, \\
  B_4&=&\frac{\langle L^4\rangle}{{\langle L^2\rangle}^2}-3
\end{eqnarray}%
The critical point can be obtained at the point where a fit to the leading
$N_s$-behavior has the least  $\chi^2$\cite{engels}.

\section{Simulations}

We perform numerical simulation using the hybrid Monte Carlo (HMC) with 
polynomial approximation\cite{odd}. We use the Wilson and DBW2\cite{DBW2} 
gauge actions.
Lattice gauge action is given by 
\be
S_g[U] = \frac{\beta}{3}((1-8c_1) \sum \mathrm{ReTr} W_{11} + c_1 \sum \mathrm{ReTr} W_{12})
\ee
where $W_{11}$ and $W_{12}$ stand for $1\times 1$ and $1\times 2$ Wilson loop.
For Wilson action the coefficient is set to $c_1=0$ and $c_1=-1.4089$
for DBW2 action.

To obtain values of observables at different $\beta$ from a single simulation, we use
the re-weighting method\cite{reweighting}. 
Statistical errors are evaluated by the Jackknife method.

\section{Results}

\begin{figure}[h]
    \resizebox{7cm}{!}{\includegraphics{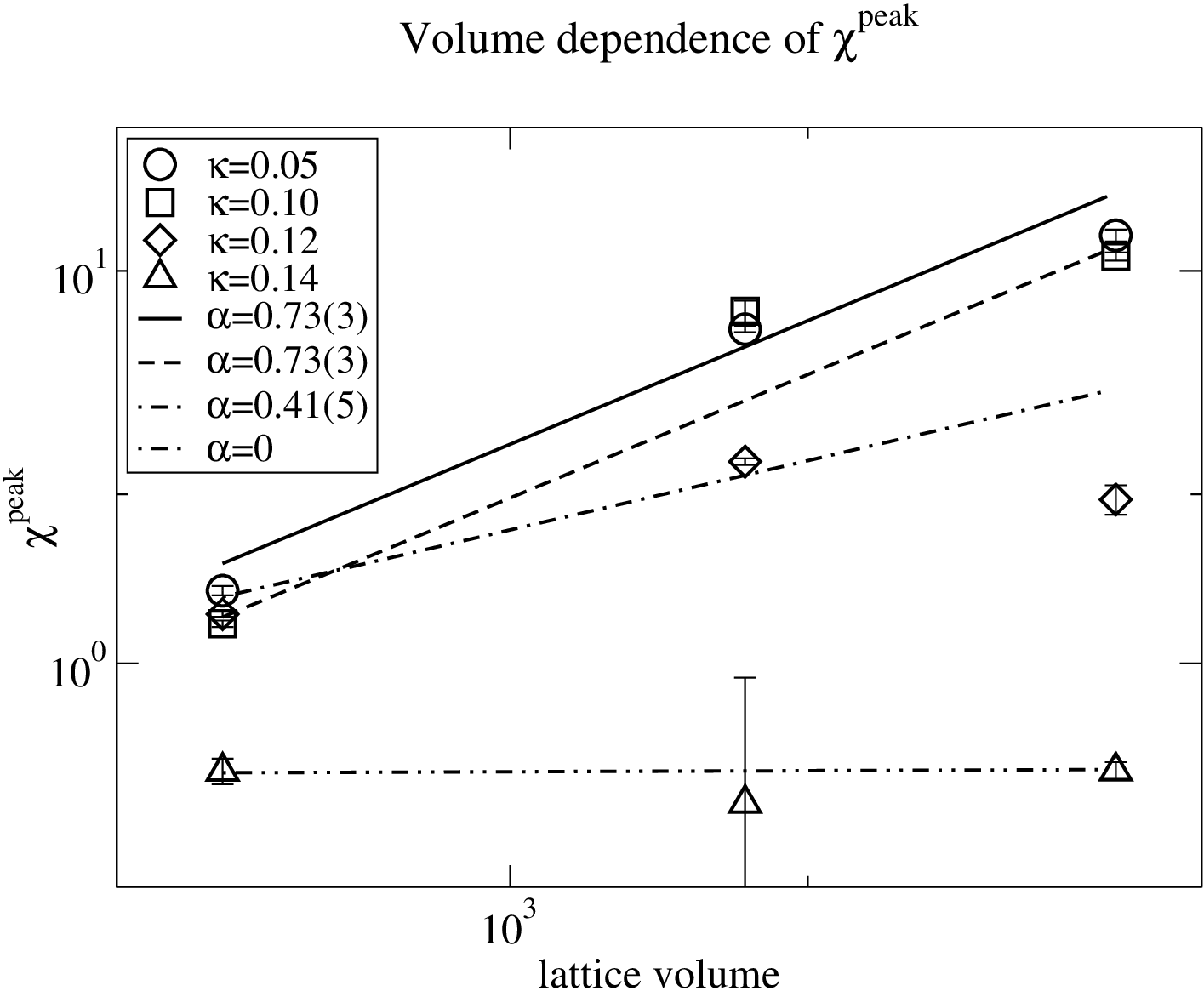}}
\vspace{-.7cm}
\caption{Volume dependence of peak of Polyakov loop susceptibility. }
\label{sus}
\vspace{-0.5cm}
\end{figure}

In Fig.\ref{sus} we show the volume dependence of peak of Polyakov loop susceptibility,
which is expected to behave as  $V^\alpha$.
The behavior is similar to the previous results\cite{multiboson}, i.e. 
$\alpha$ decreases as $\kappa$ increases. 
However the precise values of $\alpha$ are slightly different from the previous ones, 
which may indicate that our statistics is not enough to conclude.  


\begin{figure}[h]
    \resizebox{7cm}{!}{\includegraphics{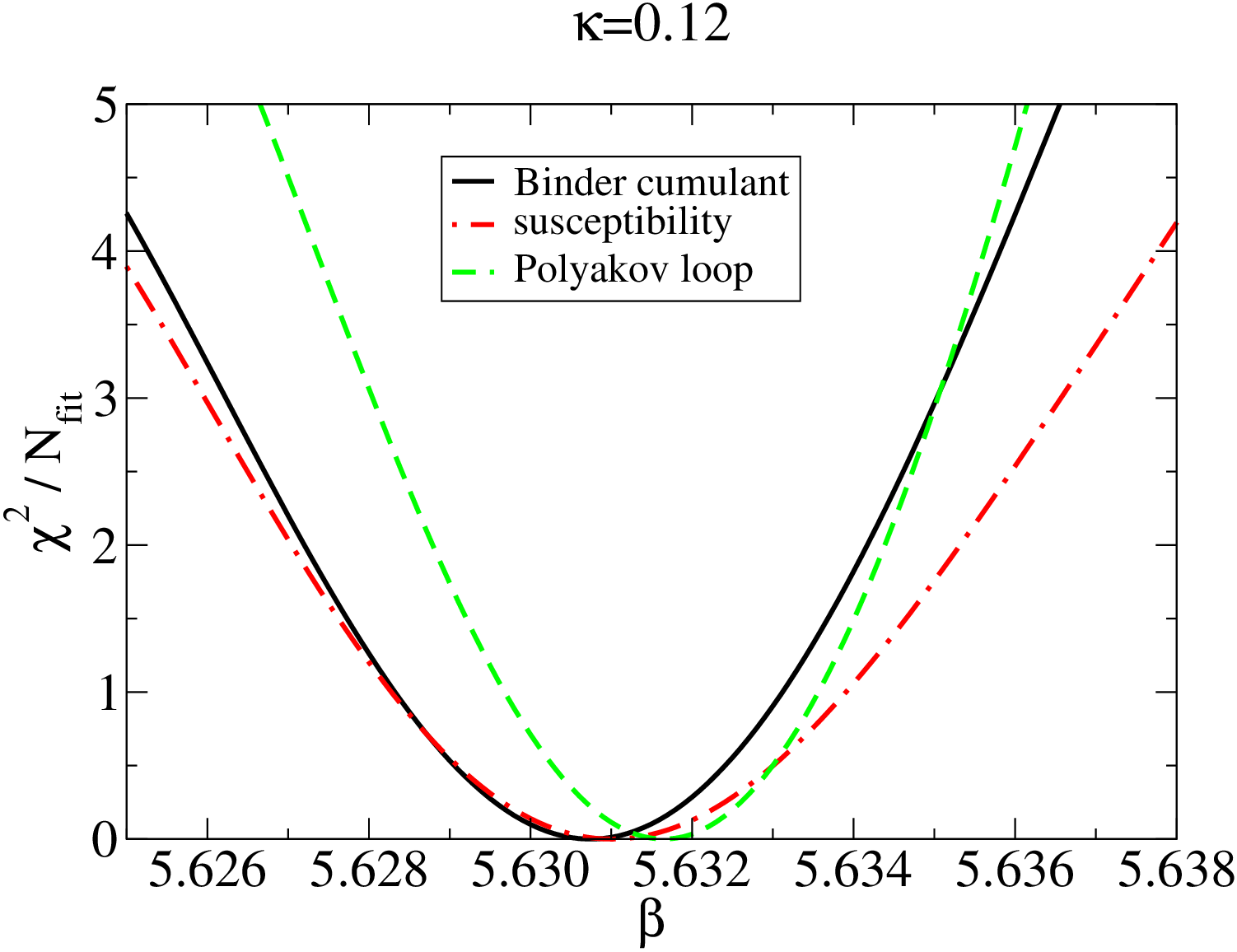}}
\vspace{-.7cm}
\caption{$\chi^2/N_{fit}$ curves at $\kappa=0.12(\beta=5.630)$. $N_{fit}=2$.}
\label{chi2}
\vspace{-0.5cm}
\end{figure}

We summarize results for $\beta_c$ from $|L|$, $\chi_v$ and $B_4$ 
in Table.\ref{table:results}.
\begin{table}[h]
\begin{center}
\begin{tabular}{c|c|c|c}
\hline
$\kappa$ & $\beta_c^{|L|}$ & $\beta_c^{\chi_v}$ &$\beta_c^{B_4}$  \\
\hline
0.05 & 5.674(4) & 5.672(3) & 5.685(6) \\
\hline
0.10 & 5.676(5) & 5.674(4) & 5.680(2) \\
\hline
0.12 & 5.630(1) & 5.631(2) & 5.631(1)  \\
\hline
0.14 & 5.595(2) & 5.592(2) & 5.616(6) \\
\hline
\end{tabular}
\end{center}
\caption{
Preliminary results of critical $\beta$ from various observables.
$\beta_c^{|L|}$, $\beta_c^{\chi_v}$ and $\beta_c^{B_4}$ stand for 
critical $\beta$ from Polyakov loop, $\chi_v$ and $B_4$ respectively. 
}
\label{table:results}
\end{table}
We estimated the critical $\beta$ using the $\chi^2$-method. 
The data were fitted to Eq.(\ref{eqn:Lfit})--
Eq.(\ref{eqn:grfit}). For Eq.(\ref{eqn:grfit}) $\omega$ was fixed to $1$ in accordance 
with \cite{engels}. 
As an example in Fig.\ref{chi2} $\chi^2/N_{fit}$ curve at $\kappa=0.12$ is shown.
Here $N_{fit}$ is the degrees of freedom for the fitting. 
We determine $\beta_c$ at the minimum point of the $\chi^2/N_{fit}$ curve.

$\beta_c^{|L|}$,  $\beta_c^{\chi_v}$ and $\beta_c^{B_4}$ should coincide 
with each other.
However  we observed that  at $\kappa=0.05, 0.10$ and $0.14$
$\beta_c^{B_4}$ is slightly different from others.  
This might be due to less statistics.
To determine  $\beta_c$ precisely  more statistics is needed. 

\begin{figure}[h]
    \resizebox{7cm}{!}{\includegraphics{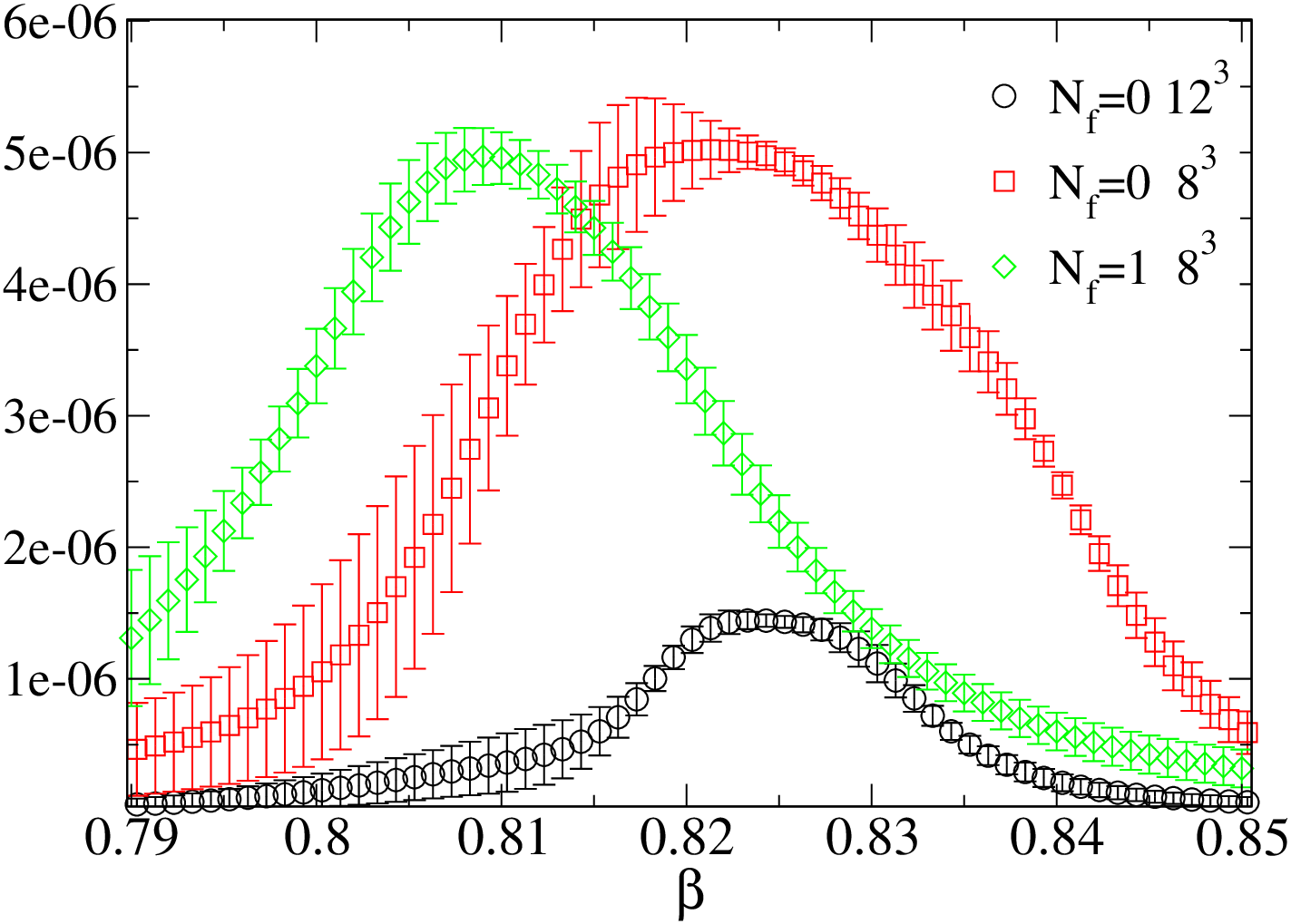}}
\vspace{-.7cm}
\caption{Plaquette susceptibility from quenched and full QCD simulations with DBW2 gauge action. 
The full QCD simulation is performed at $\kappa=0.10$.}

\label{graph:DBW2}
\vspace{-0.5cm}
\end{figure}

The results with DBW2 gauge action ( quenched and $\kappa=0.1$ ) are shown in Fig.\ref{graph:DBW2}.
We are currently performing simulations at other $\kappa$.

\section*{Acknowledgements}
The simulations were performed on NEC SX-5 at RCNP, Osaka University and
on SR8000 at Hiroshima University.

\end{document}